\documentclass[a4paper]{article}
\usepackage[affil-it]{authblk}
\usepackage{amssymb}
\usepackage{enumerate}
\usepackage{color}
\usepackage{amsmath}
\usepackage{graphicx}
\usepackage{url}

\begin{document}

% first the title is needed
\title{Author--topic profiles for academic search}

% the name(s) of the author(s) follow(s) next
%
% NB: Chinese authors should write their first names(s) in front of
% their surnames. This ensures that the names appear correctly in
% the running heads and the author index.
%

\author{Suzan Verberne\textsuperscript{1}, Arjen P. de Vries\textsuperscript{2}, Wessel Kraaij\textsuperscript{1,3}
  \thanks{\texttt{s.verberne@liacs.leidenuniv.nl}; Corresponding author}}
\affil{1. Leiden Institute for Advanced Computer Science, Leiden University\\
2. Institute for Computing and Information Sciences, Radboud University\\
3. TNO, the Netherlands}

\maketitle

\begin{abstract}

%1. What did we do
We implemented and evaluated a two-stage retrieval method for personalized academic search in which the initial search results are re-ranked using an author--topic profile. 
%2. Why did we do it
In academic search tasks, the user's own data can help optimizing the ranking of search results to match the searcher's specific individual needs. 
%3. How did we do it
The author--topic profile consists of topic-specific terms, stored in a graph. We re-rank the top-1000 retrieved documents using ten features that represent the similarity between the document and the author--topic graph.
%4. What did we find
We found that the re-ranking gives a small but significant improvement over the reproduced best method from the literature.
%5. What do we think this means
Storing the profile as a graph has a number of advantages: it is flexible with respect to node and relation types; it is a visualization of knowledge that is interpretable by the user, and it offers the possibility to view relational characteristics of individual nodes. 
%\keywords{academic search, user profiling, personalization, learning-to-rank}
\end{abstract}

\section{Introduction}

Clickthrough data plays an important role in ranking search results, especially for general-interest, high-frequency queries. %: the more a search result is clicked on by users, the higher it will be ranked in new searches~\cite{joachims2002optimizing}. 
But when the information searched becomes more specific, %(cf. the `long tail of information needs'~\cite{spink2001searching})
popularity becomes less important and less useful as a ranking criterion. Less important because for highly specific queries, relevance is more searcher-specific than for general-domain search; less useful because the amount of click data available from other users is limited. An alternative for search tasks addressing highly specialized topics is to employ the user's own data for result ranking~\cite{shen2005implicit,verberne2014query}. %for optimizing the ranking of search results. %We argue that in niche domains, relevance ranking should be inherently personalized.

Academic search is a key example of domain-specific search~\cite{hemminger2007information}. Although academic search is generally defined as a recall-oriented task~\cite{kim2011automatic}, precision is also an issue due to ambiguity of search queries -- query terms commonly have different meanings across scientific domains. Consider for example the term `search behaviour', which can refer to (according to Google Scholar): prey search behaviour, job search behaviour, the search behaviour of soccer goalkeepers, and of course information search behaviour.

A large body of previous work exists on user profiling in Information Retrieval~\cite{ghorab2013personalised}, but relatively little work addresses personalized academic search~\cite{salehi2015examining}. Approaches to user profiling and personalization often incorporate ontological information from a reference ontology~\cite{daoud2009towards}. An alternative is to collect a set of documents that are known to be relevant to the user, and generate a user profile from those documents~\cite{tang2010combination}. %In our proposed method, the user profile consists of documents that were authored by the searcher himself, and terms from those documents. The motivation behind this idea is that a personalized academic search engine can use the searcher's own texts to learn their interests, and thereby their ranking preferences. 
In this paper, we propose and evaluate a two-stage retrieval method that uses an author--topic profile for personalization. The author--topic profile consists of topic-specific terms, stored in a graph. We chose a graph representation because it is flexible with respect to node types and relation types to be included, and because it is a visualization of knowledge that is interpretable by the user. This graph is used to re-rank (2nd stage) the top-1000 documents retrieved by a baseline ranker (1st stage). The two-stage approach particularly aims at improving search precision at the top ranks~\cite{bendersky2008re}.

%There is a large body of previous work on graph models in Information Retrieval~\cite{mihalcea2011graph,alonso2016report}. Graph models are used for representing the hyperlink structure of a (web) corpus~\cite{page1999pagerank}, the citation structure and author--document relations in a bibliographic database~\cite{chiang2013exploring,zhu2017use}, social relations in social media networks~\cite{spirin2014people} and user--tag relations in folksonomies~\cite{hotho2006information}.  There is some previous work on graph-based user models in IR, most of which incorporates ontological information from a reference ontology such as the ODP~\cite{daoud2009towards,daoud2010personalized,daoud2011personalized}. Graph-based user models have been utilized for query disambiguation~\cite{tanudjaja2002persona}, query expansion~\cite{zhou2012improving}  and query suggestion~\cite{leung2008personalized}.

We evaluate our methodology using the iSearch data~\cite{lykke2010isearch}, an academic document collection with extensively described topics and relevance assessments. The content of the topic fields (information need, work task context, background knowledge, ideal answer, search terms) was written by the topic owner. We use the terms occurring in these topic fields as input for the author--topic profile. We address the following research questions: \begin{enumerate}[RQ1] \item What are the properties of a graph-based author--topic profile in which nodes represent terms? \item What are the most informative features for determining the relevance of a document given the author--topic profile? \item Can the author--topic profile be successfully exploited to improve the ranking of documents in academic search? \end{enumerate}

Our contributions compared to previous work are: (1) we propose an author--topic profile for academic search without the use of external ontologies or knowledge bases; (2) we show the effectiveness of the model to capture topical content in academic search; (3) we reach a small but significant improvement over the reproduced best method previously reported for the iSearch benchmark data.

\section{Related work}
In this section, we present previous work on user profiling and academic search (Section~\ref{sec:PAS}), and on the iSearch collection that we use for our experiments (Section~\ref{isearch}). In Section~\ref{workwithisearch} we list the previous results obtained with this collection, and describe the best performing method.%on graph representations of text, because we implement our method using a graph representation.

\subsection{User profiling and academic search} \label{sec:PAS}
% include: academic search AND ((personalization AND profile) OR "user profile"),
% exclude: folksonomy, recommendation, social networks, advertising

%NR Business as usual: Amazon. com and the academic library - MK Van Ullen (2002)
%NR Strategic planning and customer intelligence in academic libraries - R Decker (2006)
%NR The personalization information retrieval engineering research based on ontology [J] - YIN Hong-li (2008)
%NR Comparison of PubMed, Scopus, web of science, and Google scholar: strengths and weaknesses - ME Falagas (2008)
%NR Web mining functions in an academic search application - J Sivaramakrishnan (2009)
%NR Academic Search Engine Optimization (aseo) Optimizing Scholarly Literature for Google Scholar & Co. - J Beel (2009)
%NR General-purpose digital library content laboratory systems - P Manghi (2010)

%The 1999 paper by Adar et al.\cite{adar1999haystack} is probably the oldest paper that suggest a graph-based user model for information retrieval. Their model, called `haystack', represents the relationship between an individual user and the retrieval corpus; user interactions are stored in the model and used to personalized the retrieval process. 

Approaches to user profiling and personalization often incorporate ontological information from a reference ontology~\cite{daoud2009towards,pretschner1999ontology,speretta2005personalized}. %User data such as previous queries or snippets of visited web pages are automatically mapped to the terminology of the ontology. %The resulting profiles are in most studies used to re-rank the search results. %Alternatively, in the work by \cite{ma2007interest}, the ontology-based approach is exploited to improve a categorization-based retrieval system for knowledge workers. In a series of publications Daoud et al. have shown the effectiveness of personalized graph-based document ranking using a semantic user profile~\cite{daoud2009towards,daoud2010personalized,daoud2011personalized}. 
% [[hoe is de ontology nu anders van de categorization-based? en hoe relateert het categorization-based nu aan de user profiling method met topical categories die je iets verderop noemt?]] 
%In \cite{pretschner1999ontology}, an improvement of 8\% in terms of 11-point precision is reported as an effect of adding user profiles to result ranking. \cite{speretta2005personalized} report that the average rank of the results that were selected as relevant by the user improves with 33\% by adding profiles that were extracted from query history. 
An alternative option for user profile learning is term extraction: extracting prominent terms from a set of known-to-be-relevant documents~\cite{tang2010combination}. These terms can then be used for re-ranking search results based on the similarity between the user profile and the retrieved documents (e.g.~\cite{micarelli2004anatomy}). Other works use the term profile for query disambiguation~\cite{tanudjaja2002persona}, query expansion~\cite{chen1998webmate,zhou2012improving} or query suggestion~\cite{leung2008personalized,verberne2014query}. %In the paper by Chen and Sycara~\cite{chen1998webmate}, the query is expanded with the terms from the user profile that are the most correlated to the terms in the query. Kim and Chan~\cite{kim2003learning} learn a user interest hierarchy (a domain-specific taxonomy) from a set of web pages visited by a user. Each web page can then be assigned to nodes in the hierarchy for further processing in learning and predicting interests. 
%Tang et al.~\cite{tang2010combination} follow a two-step approach to web user profiling: first, a classifier identifies relevant documents, and then a user interest model is generated from these documents. 
All works report an improvement of personalization over the non-personalized baseline.

In the context of academic search, personalized retrieval has in some cases been implemented as a discovery or recommendation task~\cite{sadeh2013optimizing}. In a study addressing the effect of personalized search in a non-academic search engine for academic purposes, Salehi et al.~\cite{salehi2015examining} found that the personalization strategies of a non-academic search engine (e.g. presenting local results) did not help the academic search process, and even hampered it. This suggests that personalization strategies in the academic context should be different from personalization in ad hoc web search. In this paper, we follow up on the line of research into user profiling  through term extraction (discussed above). Our work distinguishes itself from previous work in (a) addressing academic information needs, and (b) using the searcher's own written texts, as opposed to visited or bookmarked documents.

\subsection{The iSearch collection} \label{isearch}

We use the iSearch data \cite{lykke2010isearch} for our experiments. The collection consists of 65 natural search tasks (topics) from 23 scholars from university departments of physics. The search task description form had five fields that the searchers filled in before they started to search for answers: \emph{(a) What are you looking for? (information need); (b) Why are you looking for this? (work task context); (c) What is your background knowledge of this topic? (d) What should an ideal answer contain to solve your problem or task? (e) Which central search terms would you use to express your situation and information need?} In our experiments we use these fields to populate the author--topic profile. The fields (a--d) contain roughly between 10 and 40 words each; field e is much shorter, listing between of 3 and 10 search terms.

A collection of 18K book records, 144K full text articles and 291K metadata records from the physics field is distributed together with the topics. For each topic, the developers used Indri to collect a pool of up to 200 documents from the collection and the topic owner assessed these documents on their relevance for the topic. Relevance assessments were made on a 4-point scale. %: highly relevant (3), fairly relevant (2), marginally relevant (1) and non-relevant (0). 
Normalized Discounted Dumulative gain (nDCG) is the evaluation metric commonly used for this collection, because of the graded relevance assessments in the data. 

\subsection{Previous work with the iSearch collection}\label{workwithisearch}
We build on previous work with the iSearch collection. We try to reproduce the best retrieval result and use that as baseline for our experiments. In order to create an overview of previous retrieval results obtained with the collection, we first collected the set of papers that cite the iSearch test collection paper by Lykke et al.~\cite{lykke2010isearch} (52 papers in total, according to Google Scholar, summer of 2017) and then we selected the 14 papers from that set that contain the terms `isearch'  and `ndcg'. We excluded papers that do not report results for the iSearch collection, papers that do not report retrieval results (but results for another task, e.g. term extraction), and papers that do not report nDCG scores for the data set as a whole. After excluding those, 8 papers are left. A summary of the used methods and the highest nDCG result from each paper is listed in Table~\ref{tab:prevwork}. All previous work has used Indri as index and retrieval engine.

\begin{table}[t]
\caption{Results previously obtained with the iSearch data set, sorted by nDCG scores}
{\scriptsize
\begin{tabular}{p{2.3cm}p{1cm}p{9cm}}
\hline
Paper & Best nDCG & Method\\
\hline
\cite{lioma2011sense} (SIGIR 2011) & 0.2161 &  Jelinek-Mercer smoothing and sense disambiguation; fields a and e \\
\cite{larsen2012preliminary} (SIGIR 2012) & 0.2777 &  Dirichlet smoothing; pseudo-relevance feedback, boosted technical terms\\
\cite{norozi2012contextualization} (TBAS 2012) & 0.2890 & LM; re-scoring based on inlinks and outlinks \\
\cite{dabrowska2014utilizing,dabrowska2015exploiting} (BIR 2015) &  0.3127 &  Dirichlet smoothing (optimized $\mu$), citation context\\
\cite{zhao2014language} (BIR 2014) & 0.3134 &  Jelinek-Mercer smoothing (optimized $\lambda=0.7$, stemming, stopping); field e (search terms); topics without relevant docs excluded\\
\cite{sorensen2012exploration} (TBAS 2012) & 0.3268 &  Jelinek-Mercer smoothing (optimized $\lambda=0.5$, stemming, stopping); field e (search terms)\\
%\cite{ingwersen2010does} () & 0.33 & \\
\cite{lioma2012preliminary} (IIiX 2012) & 0.3572 &  Jelinek-Mercer smoothing (optimized $\lambda$); field e (search terms)\\
\hline
\end{tabular}}
\label{tab:prevwork}
\end{table}

The best performing method is the baseline method from the paper by Lioma et al.~\cite{lioma2012preliminary}, with an nDCG of 0.3572, using only field e (search terms) as query and Indri with Jelinek-Mercer smoothing for retrieval (in terms of nDCG none of the other experimental settings in the paper by Lioma et al. beats this baseline). Table~\ref{tab:prevwork} shows quite some variation in observed effectiveness for essentially the same method (Jelinek-Mercer smoothing, optimal lambda, topic field e). We will later demonstrate that this variation can be attributed to differences in pre-processing only.

%In the next Section we describe our methods for (1) reproducing the best result by Lioma et al. as baseline and (2) improving over the baseline by using a the author--topic profile built from the information in the iSearch topics.

\section{Methods}

\subsection{Baseline} \label{sec:baseline}
We used Indri %through the Pyndri interface~\cite{VanGysel2017pyndri}. 
to reproduce the best baseline results reported by Lioma et al.~\cite{lioma2012preliminary}. Lioma et al. mention 4 different levels of preprocessing, ranging from no preprocessing at all to lowercasing, punctuation removal, stopword removal and stemming. Since the paper does not report which of the four was used to obtain the best retrieval results, we created four different versions of the iSearch index. Like Lioma et al., we use the SMART stopword list from~\cite{lewis2004rcv1} and the Porter stemmer for the stopped and stemmed versions of the index and search terms. %\footnote{In Indri, if the index is stemmed when it is built, query terms are automatically stemmed when the query is run, according to \url{https://sourceforge.net/p/lemur/discussion/836442/thread/0f00c8e5/}} 
For Jelinek-Mercer smoothing we experimented with all values for the parameter $\lambda$ that are listed in the paper (the optimal value of $\lambda$ is not reported, nor the tune set that was used for optimization). 
We preprocessed field e from the topics into a query by lowercasing the text in the field and removing punctuation. We then concatenated the tokens in one query with the Indri operator \texttt{\#combine}. We queried the iSearch collection with IndriRunQuery and retrieved the top 1000 documents (maximum) per query. We report nDCG averages over all queries. 

\subsection{Topics preprocessing}
We processed the content of the topics to populate the author--topic profile. All five fields were tokenized with NLTK, lowercased, lemmatized with NLTK (WordnetLemmatizer)\footnote{\url{http://www.nltk.org/api/nltk.stem.html}} and stopwords were removed using the SMART stopword list from~\cite{lewis2004rcv1}. Noun phrase chunking was performed with NLTK (based on NLTK pos tagging and RegexpParser)\footnote{\url{https://gist.github.com/alexbowe/879414}} and multi-word noun phrases were added as additional terms to the topic field.

We trained a Word2vec model using gensim\footnote{\url{https://radimrehurek.com/gensim/models/word2vec.html}} in order to expand the author--topic profile with related terms. For training the model we used the PN and BK parts of the iSearch collection; the title and description field from each document. %\footnote{The number of documents used for training is 309,685; The remaining 704 documents could not be processed due to malformed XML.} 
The texts were again tokenized, lowercased, lemmatized, and stopwords were removed The total collection size for word2vec training is 39,7 M words. We trained a model with dimensionality=320, window size=11, minimal number of occurrences = 5.

%39,696,063

%The NLTK WordNet lemmatizer unfortunately has a low coverage, and does not lemmatize verbs: taken $\rightarrow$ taken, investigated $\rightarrow$ investigated, fabricating $\rightarrow$ fabricating, fabricated $\rightarrow$ fabricated, particles $\rightarrow$ particles. 

%Alternative: stemmer (snowbal.EnglishStemmer). This would give manipulation $\rightarrow$ manipul and fabricating $\rightarrow$ fabric, information  $\rightarrow$  inform

\subsection{Profile building} \label{sec:profilebuilding}
\begin{figure}[t]
\includegraphics[width=12cm]{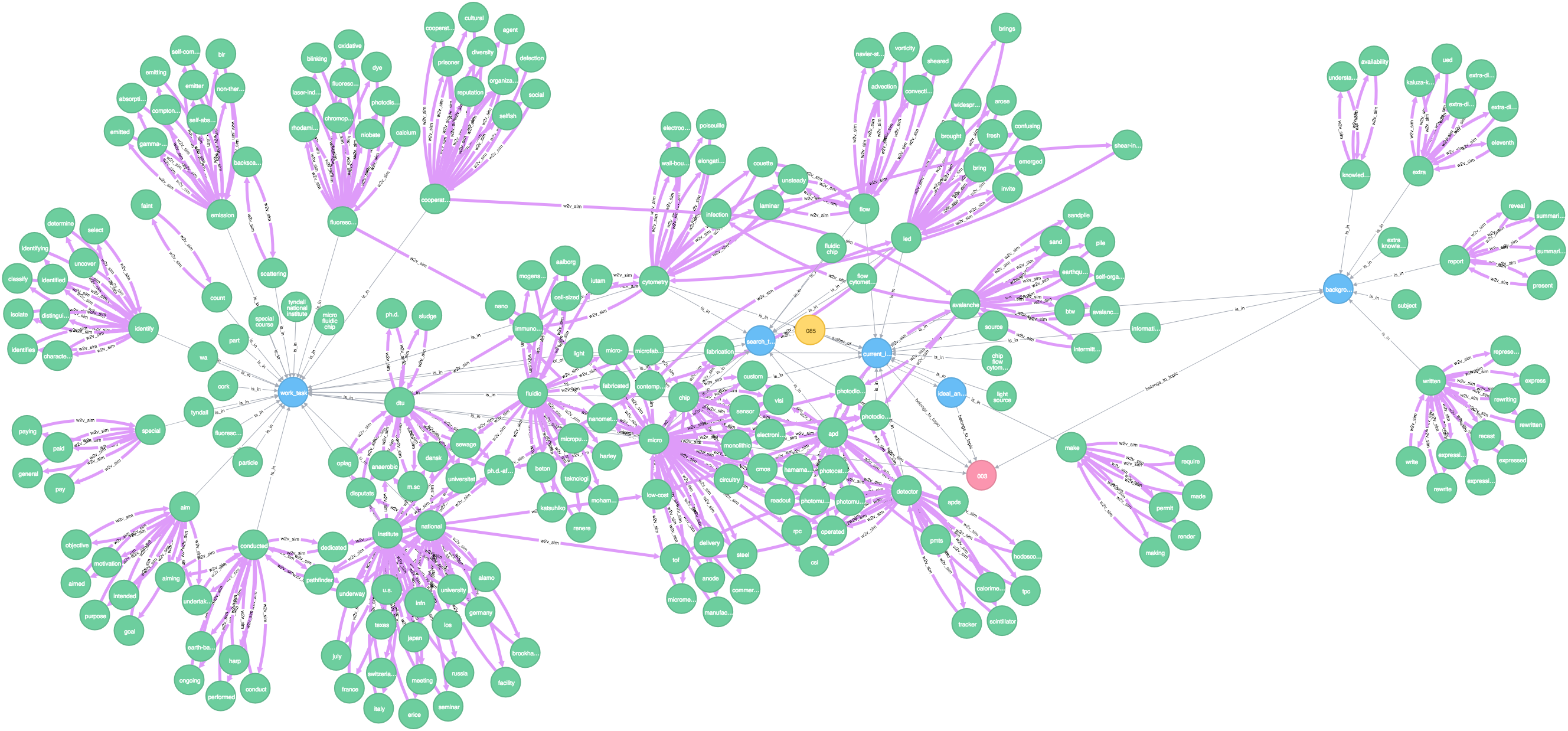}
\caption{Example author--topic graph profile. The yellow node represents the author; the red node represents the topic. The blue nodes represent the five topic fields (document nodes) and the green nodes represent the topic terms and expansion terms (term nodes).}
\label{fig:graph}
\end{figure}

We store the user profile in a graph structure for flexibility; we have future plans to expand the profile with more types of relational information, such as author and conference/journal nodes, citation relations between documents, and behavioral data such as queries and clicks on documents. %By representing the profile as a graph we allow different types of relations to be stored, such as relations between terms and documents, between queries and documents, and among documents. 
We use the graph database technology \texttt{neo4j} and its Python interface \texttt{py2neo} to build, store and access the author--topic profiles.\footnote{\url{https://neo4j.com/} and \url{http://py2neo.org/v3/}} We create a separate author--topic profile for each topic in the iSearch data. Figure~\ref{fig:graph} shows an example author--topic graph.

\paragraph{Node types and their properties} We include two types of nodes in the profile: \textsf{Document} nodes and \textsf{Term} nodes. %\textsf{Topic} nodes (with the property `name': the topic id) and \textsf{Author} nodes (with the property `name': the author id).
All words from the processed topic fields a--e were added to the graph as term nodes. We refer to these terms as \textsf{topic terms}. For each of the topic terms in the graph, we determined 10 most similar words according to the word2vec model, and added the words that have a similarity score higher than a given threshold as term nodes to the graph. %\red{Hier hebben we eigenlijk ook phrases nodig in het word2vec model. Dus zowel voor KLdiv als voor de w2v expansie. Dus misschien toch noun phrases indexeren in word2vec?}. 
These words are the \textsf{expansion terms}. %The word2vec value is added as weight to the edge between the two words. 
For the threshold $t$ we compare four different values: $\{0.1,0.3,0.5,0.7\}$.

Term nodes have three properties: `word' -- the string of the word that the node represents, `frequency' -- the total number of occurrences in the author--topic profile (word2vec expansion terms have a frequency of 0), and `kldiv' -- the importance of the term for the topic, measured using Kullback-Leibler Divergence. This term relevance metric is based on the keyphrase extraction method by~\cite{tomokiyo2003language}; we used the implementation by~\cite{verberne2016evaluation}.\footnote{\url{https://github.com/suzanv/termprofiling}} It computes the rate between the probability of the term in a foreground corpus and the probability of the term in a background corpus. In this case, the foreground corpus is compiled of the content of fields a--e of the current topic and the background corpus is the  collection of PN and BK documents that was also used for training the word2vec model. % We approximate the background frequency of multi-word terms by the frequency of the least frequent word in the term.

\paragraph{Relations and relation weights} We added the following two types of relations as edges to the graph: relations between a document node and a term node, and relations between two term nodes. Term--document relations are directed edges $T\rightarrow D$, labeled \textsf{is\_in}, with as relation weight the tf-idf score for the word in the document (where the idf is based on the iSearch corpus, using the Indri reversed index). Relations between two term nodes represent the word2vec similarity between two words. We add the edge $T_1 \longleftrightarrow T_2$ if $T_1$ is an expansion term of $T_2$ ($T_1$ might also be a topic term) and if the similarity between the two words is above the threshold $t$.

\subsection{Personalized ranking using the author--topic profile}
We retrieved at most 1000 documents per topic using the baseline method (see~\ref{sec:baseline}). The following paragraphs describe how we processed the retrieved documents, extracted features and re-ranked the documents.

\paragraph{Preprocessing the retrieved documents}
%Two options:
%\begin{enumerate}[a.] \item Use the dumpindex dv function of Indri: we get for each document the document vector: the stemmed words (stopwords replaced by [OOV]) with their position in the document. No preprocessing needed, but we should use the same stemming on the topic texts and the word2vec corpus (porter stemmer) and cannot use lemmatization or NP chunking. \item 

We used the \texttt{dumpindex dt} function of Indri to get for each retrieved document the full text. We then preprocessed all documents in the same way as the topic fields. %: tokenized with NLTK, lowercased, lemmatized with NLTK (WordnetLemmatizer) and stopwords were removed using the SMART stopword list. Words that contain numbers or punctuation marks other than - or ' were removed. 
Of the pre-processed text, we took the first 200 words for measuring the relevance to the author--topic profile. This aids the process in three ways: (1) it reduces processing time (the time needed to process a document is linearly dependent on the number of words in the document); (2) it makes the document lengths between the retrieved documents more similar; (3) it reduces the amount of noise for pdf documents (the most important content is expected in the beginning of the article, where the abstract is). %Noun phrase chunking was performed with NLTK for the first 200 words and multi-word noun phrases were added as additional terms to the pre-proecessed document (lowercased, but not lemmatized). 

%\paragraph{Connecting the retrieved documents to the author--topic profile}
\paragraph{Feature extraction}
For each retrieved document, we created a graph representation that combines the author profile with the candidate document. In this temporary graph, `is\_in' edges are created between the newly created document node and each of the term nodes representing words in the retrieved document (this includes term nodes that represent expansion terms of topic words). Note that the author--topic profile determines the vocabulary: words from the retrieved document that are not in the author--topic profile, are not added to the graph. %The following two steps (normalization and feature extraction) were repeated for every individual document. After the features were extracted for a document, the document was removed from the graph and the next document was added. %(note: no document--document relations are created, only term--document relations).
%\paragraph{Normalizing the edge weights}\label{sec:normalization} 
%The weights on all incoming edges (`is\_in' relationships) for the document node are the tf-idf weights for the words in the document. 

The stronger the relation between the author--topic profile and a retrieved document, the more relevant the document is to the topic. We quantified the relation between a document and the profile by four different feature types: degree centrality metrics, relation weight metrics, term weight metrics, and PageRank. In Table~\ref{tab:feats} we list the ten features that we implemented. %The retrieved documents are re-ranked (relative to the baseline ranking) using these ten features. 
We standardized the feature values to their z-value using the mean and standard deviation per topic (feature values for documents retrieved for different topics are not comparable) with the \texttt{preprocessing.StandardScaler} function in sklearn.

\begin{table}[t]
\caption{Description of the features that we use for re-ranking the retrieved documents}
\scriptsize{
\begin{tabular}{p{12cm}} 
\hline
(1) the baseline score, returned by Indri for the query--document combination\\
\hline
Degree centrality metrics, based on the number of connections (edges) between the document and the profile:\\
(2) absolute degree: the number of incoming edges \\
(3) relative degree: the number of incoming edges divided by number of unique terms in the document.\\
\hline
Relation weight metrics, based on the relation weights on incoming edges (tf-idf weights for the document)\\
(4) summed relation weight\\
(5) maximum relation weight\\
(6) average relation weight\\
\hline
Term weight metrics, based on the weights on the term nodes that are connected to the incoming edges to the document (KLdiv weights for the importance of the term in the profile): \\
(7) summed term weight\\
(8) maximum term weight \\
(9) average term weight \\
\hline
(10) the PageRank score for the document. We used the implementation from~\cite{mihalcea2004textrank}.* \\ %https://github.com/ashkonf/PageRank/blob/master/TextRank/textrank.py 
\hline
$\ast$ Available on \url{https://github.com/ashkonf/PageRank}\\
\end{tabular}}
\label{tab:feats}
\end{table}

%Centrality for weighted graphs. The problem of centrality is that it favors longer documents (documents with more words have more edges). Alternative: use node strength (kldiv) together with edge strength (no of occurrences), and normalize for doclength? If we simply divide the number of connections to the number of words in the document, pdf files are penalized because they are much longer.

%\paragraph{Feature selection}
%We first investigate which of the features are informative for the ranking.

\paragraph{Learning to rank}\label{sec:optimization}
We addressed the learning-to-rank problem in two different ways: (1) As a regression problem, using the graded relevance judgments (0--3) in the iSearch data as target values in the regression. We experimented with two different regressors: Linear Regression and Gradient Boosting Regression Trees (GBRT)~\cite{prettenhofer2014gradient} in sklearn. After the regressor has outputted a prediction score per document in the test partition, the documents are ranked per topic by this score and this ranking is evaluated. (2) As a rank-learning problem, using LambdaMART (as implemented in pyltr).\footnote{\url{https://github.com/jma127/pyltr}} %We should note that the evaluation module in pyltr (metric.calc\_mean) is not valid for our data because pyltr only has access to the relass for the retrieved documents. As a result iDCG is underestimated in pyltr, and nDCG is overestimated.} 
We optimized LambdaMART for nDCG@1000.

For cross validation we divided the 23 authors in 5 partitions in such a way that each partition has the same number of topics: 13. %In order to prevent overfitting we kept the topics by the same author together in the same partition. 
In five runs, we used three partitions to train the combination of features, one for tuning the hyperparameters and one for testing. We tuned the hyperparameters of GBRT and LambdaMART (max\_features, learning\_rate, max\_depth, min\_samples\_leaf) using the grid suggested by the developers of GBRT~\cite{prettenhofer2014gradient}.\footnote{\url{http://orbi.ulg.ac.be/bitstream/2268/163521/1/slides.pdf}} We set 
n\_estimators=100 (the default value) because a higher number slowed down the learning process without improving the results. 
To each test fold, we applied the optimal hyperparameter setting from the corresponding tune fold. We report average evaluation scores over all topics in all test partitions.

%The hyperparameters we experimented with:
% 1. during feature extraction (create multiple versions of the feature matrices)
%# - threshold on Word2vec similarity: include an edge between two words if the similarity > threshold. similarity_threshold={0.1,0.3,0.5,0.7}
%# - PageRank parameters: rsp=0.15, epsilon=0.00001, maxIterations=1000

\section{Results}

A challenge for the comparison of methods is that the pool of relevance assessments in the iSearch data is incomplete. Averaged over topics, 92\% of the retrieved documents are unassessed, and 64\% of the documents in the top-10 are unassessed. These will all be marked as non-relevant in the evaluation~\cite{webber2009score}. %Assuming unassessed documents to be irrelevant, then, is biased against new systems, while condensing runs by excising unassessed documents is biased in favor of new systems. Nor is the degree of bias fixed. Rather, it depends on the comprehensiveness of the pool, and therefore on the number, quality, and variety of the pooled systems.
We therefore report the results in terms of bpref in addition to nDCG. The bpref metric was designed for situations where relevance judgments far from complete~\cite{bpref2007}.% R-precision is defined as the precision after R documents are retrieved where R is the number of relevant documents for the given topic. [...] Our motivation for defining a preference-based measure is to find a measure that is robust in the face of incomplete relevance information rather than to exploit a different kind of judgment. The idea is to measure the effectiveness of a system on the basis of judged documents only. Since the scores for R-precision, MAP, and P(10) are completely de- termined by the ranks of the relevant documents in the result set, these measures make no distinction in pooled collections between documents that are explicitly judged as nonrelevant and documents that are assumed to be nonrelevant because they are unjudged. In contrast, our proposed preference measure is a function of the number of times judged non- relevant documents are retrieved before relevant documents.

\subsection{Baseline results}

The highest nDCG that we got in our attempts of reproducing the best result obtained by Lioma et al. (Section~\ref{sec:baseline}) is 0.3397 (bpref=0.3254); with stemming and stopping, and Jelinek-Mercer smoothing with optimized $\lambda=0.6$. This is lower than the reported nDCG by Lioma et al. for seemingly the same experimental setting. %And, as a comparison, S{\o}rensen et al.\cite{sorensen2012exploration} report an nDCG score of 0.3263, with the same retrieval model, $\lambda=0.5$, stemming and stopping.

The main reason why it is difficult to reproduce the exact same results, is that there are unreported details in the preprocessing of the search terms into Indri queries. The search terms field is a free text field and the topic owners used different types of separators between the terms (comma, semi-colon), there is inconsistent use of quotation marks and sometimes even formatting errors such as `cavityElectromagnetic' instead of `cavity Electromagnetic'. In particular, we noticed significant differences in nDCG scores caused by different treatment of punctuation: it matters whether hyphens are kept, and whether the symbol \texttt{'} is treated as a quote (punctuation) or an apostrophe (part of the word). For the reported result, we removed all hyphens and apostrophes, but we did not fix typos such as cavityElectromagnetic. 

In our further experiments, we used our nDCG score of 0.3397 as baseline to compare our results against. Since we are not retrieving any additional documents in the personalization step, there is an upper limit to the effectiveness of our methods. This upper bound is the evaluation for the optimal ranking of the \emph{retrieved} 1000 documents, which is nDCG=0.7177 and bpref=0.5865.

\subsection{Re-ranking results} \label{effectiveness}

Table~\ref{tab:results} shows our main results. For each method we show the result for the best performing value of the similarity threshold $t$. Overall, the best performing method is linear regression. It performs significantly better than the baseline Indri ranking ($p=0.0001$, according to a Wilcoxon signed-ranks test on the paired nDCG scores), although the effect size is small (around 10\% of the standard deviation of the baseline). The results for Gradient Boosting Regression Trees and LambdaMart both are not significantly better or worse than the baseline result ($p>0.01$). We found that the optimal values of the hyperparameters are not stable across the tune folds, which indicates sparseness of the relatively small data set. Apparently, hyperparameter tuning leads to overfitting due to the limited number of topics. % -- the optimal parameter values for the tune set are not transferrable to the corresponding test set.

\begin{table}[t]
\centering
\caption{Results in terms of nDCG and bpref. Significance testing was done with a Wilcoxon signed-ranks test on the paired scores per topic. The best result is marked in boldface; the underlined results are not significantly lower than the best result ($p>0.01$); an $\ast$ indicates a statistically significant improvement over the baseline ($p<0.01$). }
\begin{tabular}{lcc}
\hline
Method & nDCG & bpref\\
\hline
Baseline (Indri ranking) & 0.3397~ & 0.3254~	 \\
Re-ranking with Linear Regression ($t = 0.7$) & \textbf{0.3646}$\ast$ & \textbf{0.3578}$\ast$ \\
Re-ranking with Gradient Boosting Regression Trees ($t = 0.5$) &\underline{0.3515}~ & \underline{0.3406} \\
Re-ranking with LambdaMart  ($t = 0.5$) & 0.3221~ & \underline{0.3319}~\\
\hline
Upper bound given the set of retrieved documents& 0.7177~ &0.5865~\\
\hline
\end{tabular}
\label{tab:results}
\end{table}
%Fuhr 2017: Standard IR evaluation software typically prints results with four decimal digits, and so most authors copy these numbers directly into their paper. However, these four digits create the illusion of a precision that hardly ever exists. In experiments, when we have only a few hun- dred observations (e.g. relevance of documents), then four decimal places are inappropriate. So, as a minimum requirement, a single positive observation more or less in the raw data should affect at most the last digit shown.
% Effect size: In case of more complex metrics like e.g. RBP or NDCG, the difference is hard to interpret. Thus, one should apply the standard definition (https://en.wikipedia.org/wiki/Effect_size) [21, ch. 7, p. 187ff] of effect size ∆  when comparing two arithmetic means μ1 and μ2: ∆ = (μ1 − μ2)/σ (σ is the standard deviation of the baseline)

\subsection{Properties of the author--topic profile} \label{sec:graphproperties}
We expect the users' trust in personalisation of academic search to increase with transparent access to their stored profile. %Storing the profile as a graph has a number of advantages: First, as motivated in Section~\ref{sec:profilebuilding}, it is flexible with respect to other node types and relation types to be added. Second, it is a  visualization of knowledge that is interpretable by the user. Third, it offers the possibility to view relational characteristics of individual nodes. 
We explored the terminological content of the author profile by ranking the term nodes in the author graphs by three different criteria: frequency, topical importance relative to the iSearch corpus (the `kldiv' property), and degree. Frequency and topical importance do not take into account relations between terms; degree does. Table~\ref{tab:terms} shows the top-10 topic terms for one example topic (topic 001) using these criteria. The table illustrates that the three criteria select different terms. On average, each two out of the three criteria share only 1.6 terms (minimum 0, maximum 6). This indicates that the author--topic profile is a rich way of storing the content of the topic.

Note that the similarity threshold $t$ influences on the density of the graph: a lower threshold leads to more term connections being created. The mean average degree of an author--topic graph with $t=0.5$ is 2.2, while the average degree of an author--topic graph with $t=0.7$ is only 1.7. 

\begin{table}[t]
\caption{The top-10 profile terms for topic 001, using three different ranking criteria.}
\scriptsize{
\begin{tabular}{p{2.6cm}p{10cm}}
\hline
Criterion & Top-10 terms for topic 001\\
\hline
Frequency & `nano', `sphere', `peptide', `nano spheres', `manipulation', `article', `thesis', `starting', `manipulate', `immobilisation'\\
Topical importance (KLdiv) & `information', `manipulation', `manipulate', `manipulating', `scalable', `all-optical', `qip', `manipulated', `prepare', `controllable'\\
Degree ($t=0.5$) &`biomedical', `fluidic', `micro', `dielectrophoresis', `device', `chip', `gold', `au', `research', `biological'\\
\hline
\end{tabular}}
\label{tab:terms}
\end{table}

\subsection{Feature analysis} \label{sec:featimportance}
We analyzed the importance of the implemented features in two ways: by outputting the coefficients of the learned Linear Regression Model, and by evaluating rankings with single features. Both ways lead to the same four features that contribute the most to the combined ranking: the baseline score (single feature ranking nDCG is 0.3397), summed term weight (0.3187), relative degree (0.2616) and summed relation weight (0.2562). This indicates that three types of profile features provide relevant information for the personalized ranking of the documents: (1) the summed term weight (implemented as KLdiv per term) indicates the importance of document terms \emph{for the topic}, relative to the complete field of research; (2) the relative degree indicates the coverage of topic-specific terms in the retrieved document; (3) the summed relation weight (implemented as tf-idf per term) indicates the importance of the profile terms \emph{for the document}.

%Word2vec similarity: terms in the model. Many general terms: research, articles, students, activities, course, m.sc (do we want to use idf to filter them out?)]]

%Word2vec similarity: choosing the right threshold. A lower threshold leads to a more strongly connected graph.%, and costs more processing time. $t=0.5$ gives a relatively dense graph with a mean average degree\footnote{Each author graph has an average degree; we report the mean over all authors} of 2.37; $t=0.7$ gives a sparser graph with a mean average degree of 1.83. 

%\subsection{Speed/scalability}
%The time needed for building the author--topic profile is dependent on (a) the number of words in the topic documents; (b) the threshold on word2vec similarity for adding edges between similar terms. The time needed for the re-ranking is dependent on (a) the number of words in the retrieved documents, (b) also the threshold on word2vec similarity.

%Building the author--topic profile takes 8.1 seconds on average. Processing one retrieved document (reading, adding to the graph, computing features) takes xxx seconds on average. Half of this time is needed for adding the document to the graph (yy second); the other half for feature extraction (graph metrics). 

\section{Conclusion}
We implemented and evaluated a two-stage retrieval method for personalized academic search in which the initial search results are re-ranked using an author--topic profile. In this section we answer our research questions.

RQ1. ``What are the properties of a graph-based author--topic profile in which nodes represent terms?'' Storing the profile as a graph has a number of advantages. First, it is flexible with respect to other node types and relation types to be added. Second, it is a visualization of knowledge that is interpretable by the user. Third, it offers the possibility to view relational characteristics of individual nodes. We showed in Section~\ref{sec:graphproperties} that different term selection criteria (frequency, importance, and degree) select different terms, which indicates that our graph representation is a rich representation of a topic.

RQ2. ``What are the most informative features for determining the relevance of a document given the author--topic profile?'' We found in Section~\ref{sec:featimportance} that three types of profile features provide relevant information for the personalized ranking of the documents, in addition to the baseline Indri ranking: (1) The summed term importances for profile terms that occur in the document, (2) the relative degree of the document in the author--topic graph, and (3) the summed term--document relation weight for the profile terms that occur in the document. All three features represent the strength of the relation between the author--topic profile and a retrieved document, but all in a different way. The combination of these different topical relevance features constitutes the improved ranking of documents for the topic.

RQ3. ``Can the author--topic profile be successfully exploited to improve the ranking of documents in academic search?'' We found in Section~\ref{effectiveness} that re-ranking with the author--topic profile gives a small but significant improvement to the baseline Indri ranking. The best result is obtained with Linear Regression, which does not require hyperparameter tuning. Because of the small data set (65 topics), tuning leads the more complex methods to overfit on the development data.

Apart from its limited size, there are two other issues with the iSearch data set. First, the relevance assessments are incomplete (64\% of documents in the top-10 are unassessed), which makes comparison between methods more difficult~\cite{webber2009score}. Second, we noticed in reproducing the best result from the literature that (unreported) preprocessing details have a large influence on the result; better post-editing of the data would partly solve this (i.e. consistent use of punctuation marks as separators between search terms, formatting errors such as `cavityElectromagnetic'). Also, all papers should explicitly report the preprocessing that is applied to the data.

In future research, we will address the extension of the author--topic graph with other types of relational information, such as author and conference/journal nodes, citation relations between documents, and behavioral data such as queries and clicks on documents.

\bibliographystyle{splncs_srt}

\end{document}